\documentclass[fleqn,twoside]{article}
\usepackage{espcrc2,latexsym,amssymb,mathbbol}

\usepackage{graphicx}
\newcommand{\AmS}{{\protect\the\textfont2
  A\kern-.1667em\lower.5ex\hbox{M}\kern-.125emS}}

\newcommand{\be}{\begin{equation}}
\newcommand{\ee}{\end{equation}}
\newcommand{\bea}{\begin{eqnarray}}
\newcommand{\eea}{\end{eqnarray}}
\newcommand{\bml}{\begin{mathletters} \baselineskip 10pt}
\newcommand{\eml}{\baselineskip 12pt \end{mathletters}}

\newcommand{\mrm}{\mathrm}

\newcommand{\la}{\lambda}

\newcommand{\de}{\delta}

\newcommand{\bra}{\langle}
\newcommand{\ket}{\rangle}

\newcommand{\pa}{\partial}

\newcommand{\vc}[1]{\mbox{\bf #1}}

\newcommand{\tr}{\mbox{tr}}

\title{A lattice study of the Faddeev--Niemi effective action\thanks{
poster presented by L.D. at Lattice 2001
}
}

\author{L.~Dittmann\address[TPI]{Theoretisch--Physikalisches Institut, 
        Friedrich--Schiller--Universit\"at Jena, Max--Wien--Platz 1,
        07743 Jena, Germany}\thanks{lrd@tpi.uni-jena.de},
        T.~Heinzl\addressmark\address{Theoretische Physik,
        Ludwig--Maximilians--Universit\"at, Theresienstra\ss e 37, 80333
        M\"unchen, Germany}\thanks{supported in part by DFG}, 
        A.~Wipf\addressmark[TPI]}

\begin{document}

\begin{abstract}
We perform a lattice analysis of the Faddeev--Niemi effective action
conjectured to describe the low energy sector of $SU(2)$ Yang-Mills
theory. We generalize the effective action such that it contains all
operators built from a unit color vector field $n$ with $O(3)$ symmetry
and maximally four derivatives. To avoid the presence of Goldstone
bosons, we include explicit symmetry breaking terms parametrized by an
external field $h$ of mass--dimension two. We find a mass gap of the
order of 1.5 GeV.  
%The effective couplings
%are obtained via inverse Monte Carlo techniques.

\end{abstract}


\maketitle

\section{Introduction}

Recently, Faddeev and Niemi (FN) have suggested that the
infrared sector of Yang--Mills theory might be described by the
following low--energy effective action \cite{faddeev:99a},
\be
\label{FN_ACTION}
  S_{\mrm{FN}} \!  =  \!\!\! \int \!\! d^4 x \! \left[ m^2
  (\partial_\mu \vc{n})^2\! +\!  \frac{1}{4e^2} (\vc{n} \cdot
  \partial_\mu \vc{n} \times \partial_\nu \vc{n})^2 \right]\!\! .
\ee
Here, $\vc{n}$ is a unit vector field with values on $S^2$, $\vc{n}^2
\equiv n^a n^a = 1$, $ a = 1,2,3$; $m^2$ is a dimensionful and $e$  a
dimensionless coupling constant.  The FN "field strength'' is defined as
\be
  H_{\mu\nu}  \equiv \vc{n} \cdot \partial_\mu \vc{n}\times
  \partial_\nu \vc{n} \; .
\ee
FN claim that (\ref{FN_ACTION}) ``is the \textit{unique} local and
Lorentz--invariant action for the unit vector $\vc{n}$ which is at
most quadratic in time derivatives so that it admits a Hamiltonian
interpretation and involves \textit{all} such terms that are either
relevant or marginal in the infrared limit''.

It has been shown that $S_{\mrm{FN}}$ supports string--like knot
solitons \cite{faddeev:97,gladikowski:97,battye:98},
characterized by a topological charge which equals the Hopf index of
the map $\vc{n}:S^3\longrightarrow S^2$.  In analogy with the Skyrme
model, the $H^2$ term is needed for stabilization.  The knot solitons
can possibly be identified with gluonic flux tubes and are thus
conjectured to correspond to glueballs.
For a rewriting in terms of curvature free $SU(2)$ gauge fields and
the corresponding reinterpretation of $S_{{\rm FN}}$ we refer to \cite{Wipf}. 

In this contribution we are going to address the following problems:
First of all, neither the interpretation of $\vc{n}$ nor its relation
to Yang--Mills theory have been clarified. An analytic derivation of
the FN action requires an appropriate change of variables, $A \to
(\vc{n}, X )$, which decomposes the Yang--Mills potential $A$ into (a
function of) $\vc{n}$ and some remainder $X$. Although progress in
this direction has been made
\cite{langmann:99,shabanov:99a,shabanov:99b,gies:01}, there are no
conclusive results up to now.

Second, there is no reason why both operators in the FN ``Skyrme
term'', which can be rewritten as
\be
H^2 = (\pa_{\mu}\vc{n})^4 + (\pa_{\mu}\vc{n} \cdot \pa_{\nu}\vc{n})^2 \; , 
\ee
should have the same coupling. Third, and conceptually most important,
$S_{\mrm{FN}}$ has the same spontaneous symmetry breaking pattern as
the non-linear $\sigma$-model, $SU(2)\to U(1)$. Hence, it should admit
two Goldstone bosons and one expects to find \textit{no} mass gap. 
%We thank P.~v.~Baal for discussion on that point.

We have scrutinized the FN action using lattice methods. To this end
we made a sufficiently general ansatz for an $\vc{n}$--field action
that contains (\ref{FN_ACTION}) as a special case. In particular, we
allow for explicit symmetry breaking terms to avoid the appearance of
Goldstone bosons. 

\newpage

\section{Method}

After generating $SU(2)$ lattice configurations using the standard
Wilson action we fix to a covariant gauge
\cite{shabanov:99b,gies:01}.   We chose the Landau gauge (LG) defined by
maximizing $\sum_{x,\mu}\tr\,^\Omega U_{\mu}(x)$ w.r.t.~the gauge
transformation $\Omega$, leaving a residual global
$SU(2)$--symmetry. The field $\vc{n}$ is then obtained via maximizing
the functional $F_\mathrm{MAG} \equiv
\sum_{x,\mu}\tr\left(\tau_3\,^gU_{\mu}(x)\tau_3\,^gU_{\mu}(x)\right)$
of the maximally Abelian gauge (MAG) \cite{thooft,wiese}. This yields
a gauge transformation $g$ which we use to define our $\vc{n}$--field,
\bea
\label{NDEF}
\vc{n}(x) = g^\dag(x)\tau_3g(x)\ .
\eea
It is important to note that this definition leaves a residual local $U(1)$
unfixed. 

Since the configurations generated originally are randomly distributed
along their orbits, the gauge fixing is absolutely crucial for
rendering the definition (\ref{NDEF}) almost gauge invariant \cite{deforcrand}.

Our ansatz for the effective action is
$S_{\rm{eff}}=\sum_i\la_iS_i[\vc{n}]$  with couplings $\la_i$ and
operators $S_i$. Up to fourth order in a gradient
expansion there are the symmetric terms
\bea
\label{SYMM}
(\pa_{\mu}\vc{n})^2 \; ,\ (\square\vc{n})^2 \; ,\ (\pa_{\mu}\vc{n})^4
\; ,\ (\pa_{\mu}\vc{n} \cdot \pa_{\nu}\vc{n})^2\ ,
\eea
and the symmetry  breaking terms including a ``source field'' $\vc{h}$,
\bea
\label{NONSYMM}
\vc{n}\cdot\vc{h} \; ,\ (\vc{n}\cdot\vc{h})^2 \; ,\
(\pa_{\mu}\vc{n})^2\vc{n}\cdot\vc{h} \ .
\eea

The couplings $\la_i$ can be obtained by use of an inverse Monte Carlo
method \cite{Parisi}, where the (broken) Ward identities for rotational
symmetry provide an overdetermined linear system,
\be
\label{IMC}
  \sum_j \bra F_i^{ab}[\vc{n}] S^j_{,b}[\vc{n},\vc{h}] \ket \la_j = \bra
  I^a_i[\vc{n}] \ket \; .
\ee
Here, $F_i^{ab}$ and $I^a_i$ are known functions of $\vc{n}$, typically
linear combinations of n-point functions.

\begin{figure}[ht]
\includegraphics[width=0.469\textwidth]{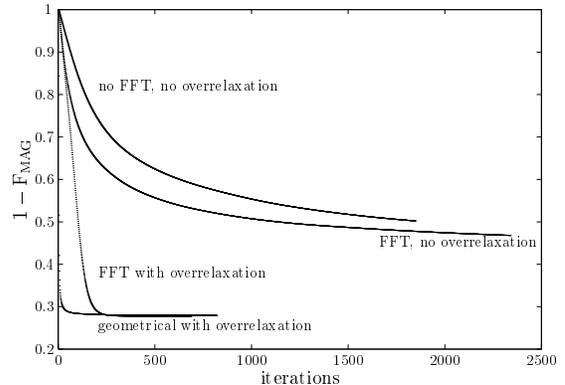}
\vspace*{-15mm}\caption{Behavior of the MAG--functional using
different algorithms.}\label{Fig.MAGS}
\end{figure}

All computations have been done on a $16^4$--lattice with Wilson coupling
$\beta=2.35$, lattice spacing $0.13\ {\rm fm}$ and periodic boundary
conditions.  For the LG we used Fourier accelerated steepest descent
\cite{Davis}. The MAG was achieved using two independent
algorithms, one (AI) being based on 'geometrical' iteration \cite{Bali:privat},
the other (AII) analogous to LG fixing (see Fig.~\ref{Fig.MAGS}).

\section{Results}

\begin{figure}[ht]
  \includegraphics[width=0.469\textwidth]{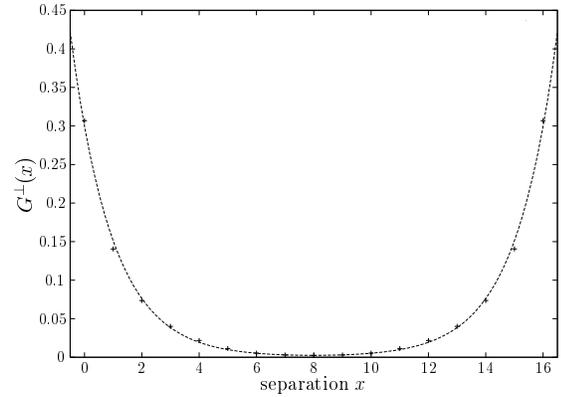}
   \vspace*{-15mm}\caption{The transverse correlation function
$G^\perp$, fitted to $G^\perp (x) \sim \cosh(-M(x-L/2))$.}
  \label{Fig.coshfitt}
\end{figure}

As expected, we observe a non-vanishing expectation value of the
field (one-point function) in the three-direction that can be thought
of as a 'magnetization' $\mathfrak{M}$, $ \bra n^a \ket = \mathfrak{M} 
\, \de^{a3}$. Thus, the global symmetry is broken explicitely according to the
pattern $SU(2)\to U(1)$. This also shows up in the behavior of the
two--point functions (Fig.~\ref{Fig.twopointfunctions}), which exhibit
clustering, $\bra n^3(0) n^3(x) \ket \sim  \bra n^3 \ket \bra n^3 \ket =
\mathfrak{M}^2$, for large distances. Furthermore, the transverse
correlation function (of the would-be Goldstone bosons)
\vspace*{-1mm}
\bea
G^\perp(x) \equiv \frac{1}{2} \bra n^i(0)n^i(x) \ket,\ i=1,2\ ,
\eea
decays exponentially as shown in Fig.~\ref{Fig.coshfitt}. This means
that there is a nonvanishing mass gap $M$ whose value can be obtained
by a fit to a $\cosh$--function. 

The numerical values of the observables, $\mathfrak{M}$, $M$ and the
transverse susceptibility, $\chi^\perp \equiv \sum_x G^\perp(x)$, are
summarized in Table 1 for both algorithms:
\begin{figure}[ht]
  \includegraphics[width=0.469\textwidth]{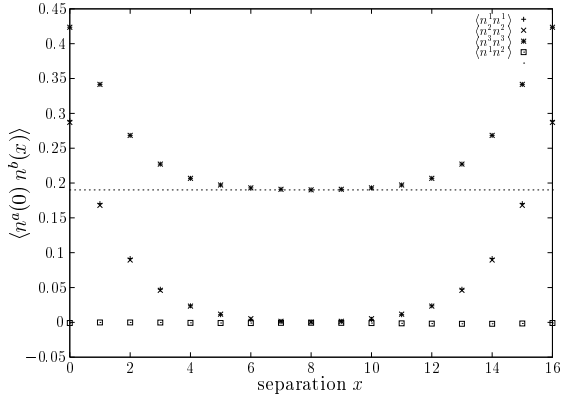}
  \vspace*{-15mm} \caption{Two-point correlators of the field $\vc{n}$
  obtained via algorithm AI. The dotted line represents the (squared)
  VEV of $\vc{n}$, $\bra n^3 \ket^2 = \mathfrak{M}^2.$ The same behavior
  is obtained via AII with slightly different plateau value (see
  Table~1).}
  \label{Fig.twopointfunctions}
\end{figure}
\begin{table}[ht]
\caption{Numerical values for some observables (all numbers in units
of the lattice spacing).}
\label{table:1}
\newcommand{\hp}{\hphantom{$-$}}
\newcommand{\cc}[1]{\multicolumn{1}{c}{#1}}
\renewcommand{\tabcolsep}{0.99pc} % enlarge column spacing
\renewcommand{\arraystretch}{1.2} % enlarge line spacing
\begin{tabular}{@{}lllll}
\hline
   & {$\mathfrak{M}$} & {$\chi^\perp$} & $M$ & {$\chi^\perp M^2$} \\
\hline
AI & 0.436 & 0.636 & 0.95 & 0.53\\  
AII& 0.352 & 0.596 & 1.01 & 0.58\\\hline
\end{tabular}%\\[2pt]
\end{table}
The slight disagreement between AI and AII is expected from our
still somewhat low statistics. The numerical results for the mass gap
$M$ lead to a value of about 1.5 GeV in physical units.  The last
column is a measure for the accuracy of the \emph{minimal} ansatz
consisting of the first (leading) terms of (\ref{SYMM}) and
(\ref{NONSYMM}), respectively.
%$S_{\rm{eff}} = \la_1(\pa_{\mu}\vc{n})^2 + \la_2\vc{n}\cdot\vc{h}/|\vc{h}|,\
%\la_2 = \la_1|\vc{h}|$
In this case the (continuum) mass gap is determined by the ``source'', $M^2=|\vc{h}|$. In addition, one has the exact
Ward identity $\mathfrak{M}=\chi^\perp M^2$.  Using this relation one
obtains the rough estimate that $M\simeq 1.2 {\ \rm{GeV}}$. Compared
to the  `exact' (fitted) value of $M\simeq 1.5 {\
\rm{GeV}}$  we find a qualitative agreement already
to lowest order.

The effective couplings have to be determined by solving
(\ref{IMC}). Results already obtained will be reported elsewhere.

\section*{Acknowledgements}

The authors thank S.~Shabanov for suggesting this
investigation and P.~van Baal for raising the issue of Goldstone
bosons. L.D.~is indebted to M.~M\"uller-Preu\ss ker and G.~Bali
for their assistance. Discussions with E.~Seiler, P.~de~Forcrand and
P.~van Baal are gratefully acknowledged.

\end{document}